# Relaxation processes in one-dimensional self-gravitating many-body systems

Toshio Tsuchiya*

*National Astronomical Observatory, Mitaka, 181, Japan*

Naoteru Gouda

*Department of Earth and Space Science, Osaka University, Toyonaka, 560, Japan*

Tetsuro Konishi

*Department of Physics, Nagoya University, Nagoya, 464-01, Japan*

(May 24, 1995)

## Abstract

Though one dimensional self-gravitating $N$-body systems have been studied for three decade, the nature of relaxation was still unclear. There were inconsistent results about relaxation time; some initial state relaxed in the time scale $T \sim N t_c$, but another state did not relax even after $T \sim N^2 t_c$, where $t_c$ is the crossing time. The water-bag distribution was believed not to relax after $T \sim N^2 t_c$. In our previous paper, however, we found there are two different relaxation times in the water-bag distribution;in the faster relaxation ( microscopic relaxation ) the equipartition of energy distribution is attains but the macroscopic distribution turns into the isothermal distribution in the later relaxation (macroscopic relaxation). In this paper, we investigated the properties of the two relaxation. We found that the microscopic relaxation time is $T \sim N t_c$, and the macroscopic relaxation time is proportional to $N t_c$, thus the water-bag does relax. We can see the inconsistency about the relaxation times is resolved as that we see the two different aspect of relaxations. Further, the physical mechanisms of the relaxations are presented.

*Electronic address: tsuchiya@yso.mtk.nao.ac.jp

## I. INTRODUCTION

One-dimensional self-gravitating many-body systems have been studied for three decade, both from astrophysical and chaotic dynamical interests. In the astrophysical point of view the system is a simple model of stellar systems, such as globular clusters, elliptical galaxies, and clusters of galaxies, and is convenient to draw out basic understanding of evolution and relaxation. On the other hand, it has been a problem in statistical mechanics whether all globally coupled systems with large degrees of freedom are ergodic and usual statistical mechanics is applicable.

Hohl [1–3] first asserted that there is a relaxation time scale of $N^2 t_c$, where $N$ is the number of particles and $t_c$ is the characteristic time which is approximately the time for a member to traverse the system. In late 1980's, however, several authors show that there is no time scale of $N^2 t_c$ in relaxation, but the character of evolution depends on the initial condition. Some initial states appears to relax on the time scale $N t_c$ [4–6], which is much shorter than Hohl's prediction. On the other hand, another initial states, such as the water-bag distribution, which we study here, do not relax in a time longer than $N^2 t_c$ [7]. This inconsistency in the relaxation is still remains to be a problem.

We have studied the evolution of the water-bag initial distribution, which was a counter example of the fast relaxation, $N t_c$. In our previous paper [8] ( we refer to it as Paper I ), we showed that there exist two characteristic time scales. The shorter one is the *microscopic relaxation*, which means that mixing in energy space has developed completely as the result of mutual interaction but the energy distributions in the $\mu$ space ( one-body phase space) are not transformed into that of the thermal equilibrium. This transformation occurs in a much longer time scale, which we refer to as the *macroscopic relaxation time*.

Our aim is to understand the physical mechanism of the relaxations and then to resolve inconsistent results about the relaxation. Sec. II gives a description of our model and summarize the evolution of the water-bag distribution found in Paper I. In Sec. III dependence of the microscopic relaxation on the number of the particles is studied numerically, and physical interpretation is presented. In Sec. IV the macroscopic relaxation is considered, and conclusions and discussions are given in Sec.V

## II. MODEL

The system comprises $N$ identical mass sheets, each of uniform mass density and infinite in extent in the $(x, y)$ plane. We call the sheets *particles* in this paper. The particles are free to move along $x$ axis and accelerate as a result of their mutual gravitational attraction. The Hamiltonian of this system has a form of

$$H = \frac{m}{2} \sum_{i=1}^{N} v_i^2 + (2\pi G m^2) \sum_{i<j} |x_j - x_i|, \qquad (1)$$

where $m$, $v_i$, and $x_i$ are the mass (surface density), velocity, and position of $i$th particle respectively.

In our calculation, we employ the size of the system and the typical velocity as units of length and velocity. Then time is measured by the crossing time,

$$t_c = (4\pi G \rho_{\mathrm{av}})^{-1/2}, \tag{2}$$

where $\rho_{\mathrm{av}}$ is the mass divided by the width of the distribution at initial time.

The potential of the system with finite $N$ has no divergence, and the equi-energy hyper surface is compact. Thus we can define the thermal equilibrium by the maximum entropy state. The distribution of the thermal equilibrium is called the *isothermal distribution*. This equilibrium is the statistical equilibrium of general kind of many-body systems. Figure 1 shows the isothermal distribution in $\mu$ space.

In addition to the thermal equilibrium, self-gravitating systems have another concept of equilibrium, *dynamical equilibrium*. It is defined mathematically in the continuous limit to the infinite number of particles in the system while the total mass is constrained to be finite. A state of the system is described by a distribution function $f(x,v)$, which is the number density of the particles at $(x,v)$ in $\mu$ space. Since the mass of a particle goes to zero as the number of particles goes to infinity, the gravitational interaction between two constituents becomes negligible and the particles move by the force due to the mean potential $\Phi$, where $\Phi$ is determined by the Poisson equation,

$$\nabla^2 \Phi = 4\pi G \int f(x,v)\,dv. \tag{3}$$

Dynamics of the system is subjected by the collisionless Boltzmann equation,

$$\frac{\partial f}{\partial t} + v \frac{\partial f}{\partial x} - \nabla \Phi(x) \frac{\partial f}{\partial v} = 0. \tag{4}$$

A dynamical equilibrium is a stationary solution of the Poisson and the collisionless Boltzmann equations. There exist many dynamical equilibria [see e.g., [9]] including the *water-bag distribution*, which has a homogeneous phase density. Figure 2 shows the water-bag distribution with $N = 1024$ in the $\mu$ space.

The systems with the continuous medium approximate systems with finite but large number of particles. If the number is very large, a finite system stays in a dynamical equilibrium for a long time. However, it is not true equilibrium, and the scattering among the discrete particles transforms the state gradually from the dynamical equilibrium to the thermal equilibrium. We call this transition the macroscopic relaxation, in contrast with the microscopic relaxation, which leads the system to the equipartition of the individual energy. We summarize briefly the results of Paper I.

In Paper I, we introduce three quantities to analyze the evolution of the systems: the equipartition and the power spectrum density of fluctuation of individual particle energy, and the maximum Lyapunov exponent. The combination of these quantities revealed the microscopic evolution of the system. As the result, we can distinguish the two different time scale from the variation of the deviation $\Delta(t)$ from equipartition. The definition of $\Delta(t)$ is shown below.

The specific energy (energy per unit mass) $\varepsilon_i(t)$ of $i$th particle is given by

$$\varepsilon_i(t) = \frac{1}{2}v_i^2(t) + 2\pi Gm \sum_{j=1}^{N} |x_j(t) - x_i(t)|. \tag{5}$$

If the evolution of the system is ergodic in the $\Gamma$-space, the long time average of the specific energy takes a unique value for all $i$, i.e.

$$\overline{\varepsilon_i} \equiv \lim_{T \to \infty} \frac{1}{T} \int_0^T \varepsilon_i(t)dt = \varepsilon_0 \equiv 5E/3. \tag{6}$$

The degree of the deviation from the equipartition is measured by the quantity,

$$\Delta(t) \equiv \varepsilon_0^{-1} \sqrt{\frac{1}{N} \sum_{i=1}^{N} (\overline{\varepsilon_i}(t) - \varepsilon_0)^2}, \tag{7}$$

where $\overline{\varepsilon_i}(t)$ is the averaged value until $t$. In the numerical scheme, $\varepsilon_i(t)$s are sampled at every crossing time $t_c$, and the average is defined simply by the summation of the samples divided by the number of the samples.

The variation of $\Delta(t)$ of the system with the water-bag initial distribution for $N = 64$ is shown in Fig.3. The plateau at the beginning represents the collisionless phase, because in the collisionless phase the individual energies are conserved. After $t \sim 100$, $\Delta(t)$ begins to decrease as $t^{-1/2}$, which means that the fluctuation behaves as the same manner as the thermal noise. The transition from constant of $\Delta(t)$ to the power law, $\Delta(t) \propto t^{-1/2}$ determines the microscopic relaxation. If the water-bag distribution is the thermal-equilibrium, then no more change is expected and $\Delta(t)$ goes to zero as $t$ increases. However, it is found that $\Delta(t)$ increases at some $10^6 t_c$. At this time, the water-bag is transformed into the isothermal distribution. The reason why $\Delta(t)$ increases is that the isothermal distribution contains more high energetic particles than the water-bag distribution and then the amplitude of the fluctuation is larger. The macroscopic relaxation is estimated by the increase of $\Delta(t)$.

## III. MICROSCOPIC RELAXATION

### A. Numerical results

The microscopic relaxation is that the individual particle energies mix well then the system attains the equipartition. In Paper I we demonstrated that the diffusion of the individual energies are similar to random walk processes, which yield the power spectrum density (PSD) of the Lorentz distribution, $P(f) \propto (f^2 + \gamma^2)^{-1}$, where $\gamma$ is constant. One way to determine the time scale of microscopic relaxation is to get the frequency, $f \sim \gamma$, where the distribution changes its gradient. The reciprocal of $\gamma$ gives the relaxation time scale. We take the other way to determine the time scale that the transition of variation of $\Delta(t)$ from constant to the power of $t^{-1/2}$ gives the same relaxation time scale. The relaxation time is simply estimated by crossing of two lines in Fig. 3: one is constant $\Delta(t)$ and the other is $\Delta(t) \propto t^{-1/2}$. We examined 100 different initial states; all have the water-bag distribution macroscopically but created from different random seeds at the initial time. The all 100 initial states are different only in the microscopic distribution. We found all runs gave almost the same microscopic relaxation time. Therefore the time scale is universal for the water-bag distribution. Figure 4 shows the microscopic relaxation time for various $N$, the number of particles. We calculated $N = 64, 128, 256, 512$, and $1024$. The doted line in the figure is the line of $T = 2N\, t_c$. The result tells us that the microscopic relaxation time, $T_m$, increases linearly as $N$ increases;

$$T_m \sim N\, t_c. \tag{8}$$

### B. Physical mechanism of the relaxation

The microscopic relaxation is the process that the energy of individual particles varies from the initial value. This variation causes the mixing of the energies and leads the system to the equipartition. The Lorentz distribution of the PSD guarantees that the variation is the same as the Brownian motion. The relaxation driven by Brownian motion in a stellar system is studied by Chandrasekhar [10–12]. This analysis can be applicable to our case.

Consider a particle located near the center at rest. The gravitational acceleration of the particle is proportional to the difference between the number of particles at the left side, $N_{\text{left}}$ and the right side, $N_{\text{right}}$, of the particle,

$$\dot{v} = 2\pi G m \left[ N_{\text{right}} - N_{\text{left}} \right] \tag{9}$$

The field particles travel across the system and the number of particles in both the left and the right sides fluctuate in time around its average. If we assume the fluctuation of distribution of field particles is random and independent, the particle at the center has a random force,

$$\dot{v} \sim 2\pi G m \sqrt{N} \tag{10}$$

Since this fluctuation is created by motion of field particles, the typical lifetime of the fluctuation should be the crossing time,

$$t_c = \left[ \frac{4\pi G N m}{L} \right]^{-1/2}, \tag{11}$$

where $L$ is the length of the system.

The fluctuation of velocity, $\delta v$, caused by the random force is,

$$\delta v \sim \sqrt{\pi G m L} \tag{12}$$

and if the fluctuations of field particle distribution occur independently, statistical theory asserts that the dispersion of the velocity fluctuation increases linearly in time,

$$(\delta v)^2 \frac{t}{t_c} \tag{13}$$

Relaxation is accomplished when the dispersion grows as much as the typical value of velocity, $V$, in the system,

$$V^2 \sim 4\pi GmNL. \tag{14}$$

Thus we get a time scale,

$$t \sim N t_c \tag{15}$$

This time scale quite agrees with the numerical result ( eq.8 and Fig. 4).

## IV. MACROSCOPIC RELAXATION

The macroscopic relaxation corresponds to a phenomenon that the macroscopic distribution is transformed from that of a dynamical equilibrium to that of the thermal equilibrium. The time scale of the transition from the water-bag to the isothermal distribution is much larger than the microscopic relaxation time.

At the transition the distribution in the energy space becomes wider. The appearance of high energetic particles causes the magnification of the amplitude of the fluctuations. In order to examine this change, we introduce the locally averaged energy, which is defined by,

$$\langle \varepsilon_i \rangle(t) \equiv \frac{1}{\Delta t} \int_{t-\Delta t}^{t} \varepsilon_i(t) dt. \tag{16}$$

We took $\Delta t \sim 20000$, which is much longer than the microscopic relaxation time in the water-bag distribution. By analogy with the $\Delta(t)$, we introduce the deviation of the energy fluctuation from the value of the equipartition,

$$\delta(t) \equiv \varepsilon_0^{-1} \sqrt{\frac{1}{N} \sum_{i=1}^{N} (\langle \varepsilon_i \rangle(t) - \varepsilon_0)^2}, \tag{17}$$

which gives the time variation of the energies averaged over $\Delta t$. The sudden increase of $\Delta(t)$ in Fig. 3 corresponds to the beginning of violent variation of $\delta(t)$ at $t \sim 5 \times 10^6 \, t_c$ in Fig. 5.

In the case of the microscopic relaxation of the water-bag distributions, different microscopic distributions (created by different random seeds) yield the definite relaxation time scale. For the macroscopic relaxation, however, we found a distribution of relaxation time which has a range over an oder of magnitude. Therefore the mechanism of the macroscopic relaxation is surmised to be different from that of the microscopic relaxation. In this case, the distribution of the relaxation time is a big clue to find the mechanism.

Figure 6 shows the probability distribution of the relaxation time of the system with $N = 64$. We took 100 different initial states, statistical ensemble, as follows: As described in Paper I the initial states are realized by random distributions of particles which form uniform averaged phase density. Different seeds of the random number generator yield different random distributions microscopically but all gives the water-bag distribution macroscopically. The relaxation time of each run, $T_M$, is determined by the figure of $\delta(t)$, and $P(T_M) \, dT_M$ gives fraction of the number that relaxes in a time interval $dT_M$ around $T_M$. It is clear from the figure, that the distribution has a exponential distribution,

$$P(T_M) = \frac{1}{\langle T_M \rangle} e^{-T_M/\langle T_M \rangle}, \tag{18}$$

where,

$$\langle T_M \rangle = 2.8 \times 10^6 t_c, \quad \text{for } N = 64, \tag{19}$$

and this gives the expectation value of the relaxation time.

Before we proceed to speculate the mechanism it is useful to remind us the phase space dynamics of the system.

Any state of the system with $N$ particles can be described by a certain point in the $2N$ dimensional phase space ($\Gamma$ space). Each point yields some macroscopic distribution, such as the water-bag or the isothermal. In $\Gamma$ space, there is a region where all phase points yield the water-bag distribution. At the beginning the phase point is located in the region and then it moves in the region as the system evolves. The macroscopic distribution is the water-bag during the phase point stays in the region. When the point escapes from the region, the macroscopic distribution is transformed from the water-bag distribution into the isothermal distribution because the isothermal distribution is defined as the maximum entropy state, which means that it occupies the largest, and usually most region in the $\Gamma$ space.

Now we should explain the following facts:

1. The water-bag exhibits several properties of thermal equilibrium, such as conversion of Lyapunov exponents, and equipartition of energy, which suggest ergodicity.

2. The probability distribution of the macroscopic relaxation time is a exponential function.

One possibility to explain these facts is that there exists a barrier which is an obstacle for a phase point to go out from the water-bag region. Due to this barrier a phase point is restricted in the region for a long time, then it travels all over the region as if the water-bag region is ergodic. However, the barrier is not perfect but there is a small gate from which a phase point in the water-bag region can escape. A phase point travels in the region in a very complicate way and almost ergodic, thus the point finds the gate and escape at random. Now we show a simple model which might explain the numerical results well as follows: as the simplest case, we assume that the escape probability is uniform in the region. Suppose an ensemble of the phase points in the region. At the beginning the ensemble contains $n(0)$ points, but they escape from the region with constant rate $r$, then the number of points which stay in the region at $t$ decreases as $n(t)$. The differential equation of $n(t)$ is

$$d\,n(t) = -n(t)\,r\,dt, \tag{20}$$

then

$$n(t) = n(0)\,e^{-rt}. \tag{21}$$

The probability distribution of the macroscopic relaxation time corresponds to the derivative of $n(t)$,

$$-\frac{1}{n(0)}\frac{d\,n(t)}{dt} = r\,e^{-rt} \tag{22}$$

It is equivalent to eq.(18). Therefore this simplest model can explain the facts obtained by the simulations.

Next, in order to investigate the dependence of the time scale on the number of particle, the same procedure was applied to the system with different $N$; $N = 16$, $32$, $128$, and $512$. Figure 7 shows the results. Especially for $N = 512$, we observed that 50 runs of the maximum integration until the time $T = 10^6 t_c$ did not relax. Thus we can not determine the time of the relaxation for $N = 512$, but by combining with the exponential probability distribution, we can restrict the region that the *true* relaxation time lies probably. The arrow indicated the region of 90% confident level. These data is approximated by a linear relaxation,

$$\langle T_M \rangle = 4 \times 10^4 \, N \, t_c, \tag{23}$$

which is shown by the dashed line.

## V. CONCLUSIONS AND DISCUSSIONS

The numerical simulation of the one-dimensional self-gravitating many-body systems revealed the existence of two different kind of relaxations. One is the microscopic relaxation where the system attains the equipartition of individual energies though the macroscopic distribution shows no change. The other is the macroscopic relaxation where the macroscopic distribution is transformed into that of the thermal equilibrium.

We have shown the dependence of the microscopic and the macroscopic relaxation times on the number of the particles for the water-bag initial distributions. Our conclusions are summarized as follows:

1. For the water-bag initial distributions, the microscopic relaxation time is about $N$ times the crossing time, $T_m \sim N \, t_c$.

2. The physical mechanism of the microscopic relaxation is the diffusion of the energies of individual particles. Its driving force is the gravitational force of random fluctuation as a random force.

3. The macroscopic relaxation time, where the transition of the macroscopic distribution from the water-bag to the isothermal occurs, depends linearly on the number of the particles, $T_M \sim 4 \times 10^4 \, N \, t_c$.

4. The macroscopic relaxation time $T_M$ depends on initial conditions. It has exponential distribution as

$$P(T_M) = \frac{1}{\langle T_M \rangle} e^{-T_M/\langle T_M \rangle}. \tag{24}$$

5. From the above facts, it is inferred that there is a region in $\Gamma$ space where the macroscopic distribution yields the water-bag, and a phase point is confined to the region for a long time, then it escapes through a small gate to the isothermal region at random. This hypothesis naturally explains why the water-bag behaves as if it is the thermal equilibrium, and why the macroscopic relaxation time has the exponential distribution.

The microscopic relaxation time, $T_m \sim N t_c$, agrees with the time scale found in ref [4–6], which concerned initial dynamical phase started from dynamically non-equilibria (the virial ratio is not unity). In such cases, both microscopic and macroscopic relaxations occur at the same time scale, thus the two relaxations were believed to be equivalent. The water-bag distribution had been known as one of the counter examples of the relaxation of $N t_c$, because the water-bag did not show the transformation of the macroscopic distribution to the isothermal distribution until $N^2 t_c$ (for $N = 200$) [7,13]. We found, however, the water-bag distribution does relax to the thermal equilibrium at the time scale much longer than the previous calculations. Luwel and Severne [13] already pointed out the existence of the collisionless mixing, which is equivalent to the microscopic relaxation in our papers. We determine the time scale as a function of the number of the system, which agrees the relaxation time of the system evolved from dynamically non-equilibria. Therefore we conclude that the inconsistency of the relaxation of the water-bag distribution is due to the unawareness of the difference between the microscopic and the macroscopic relaxations.

In our hypothesis, the water-bag region is enclosed by some barrier, but what the barrier is, or how the motion of the phase point is restricted, is unknown. It could be that the water-bag is the special case. However, if these properties are general for any stable dynamical equilibria, then the similar barriered regions are present outside the water-bag region. The fact that in the system with less particle ($N \leq 20$) the phase space is divided into segments [14] supports our hypothesis. The studies of individual orbits during the quasiequiribrium of the water-bag and the transition to the isothermal will reveal the mechanism. We are working on the issue and will report it in the next paper.

## ACKNOWLEDGMENTS

The computation was carried out on Hewlett-Packard HP730 of the theoretical astrophysics division, National Astronomical Observatory.

# REFERENCES

[1] F. Hohl and D. T. Broaddus, Phys. Lett. A **25**, 713 (1967).
[2] F. Hohl and M. R. Feix, Astrophys. J. **147**, 1164 (1967).
[3] F. Hohl and J. W. Campbell, Astron. J. **73**, 611 (1968).
[4] M. Luwel, G. Severne, and P. J. Rousseeuw, Astrophys. Space Sci. **100**, 261 (1984).
[5] G. Severne, M. Luwel, and P. J. Rousseeuw, Astron. Astrophys. **138**, 365 (1984).
[6] C. J. Reidl, Jr. and B. N. Miller, Phys. Rev. A **46**, 837 (1992).
[7] H. L. Wright, B. N. Miller, and W. E. Stein, Astrophys. Space Sci. **84**, 421 (1982).
[8] T. Tsuchiya, T. Konishi, and N. Gouda, Phys. Rev. E **50**, 2706 (1994).
[9] J. Binney and S. Tremaine, *Galactic Dynamics* (Princeton Univ. Press, Princeton, 1987).
[10] S. Chandrasekhar, Astrophys. J. **93**, 285 (1941).
[11] R. E. Williamson and S. Chandrasekhar, Astrophys. J. **93**, 305 (1941).
[12] S. Chandrasekhar, Astrophys. J. **93**, 323 (1941).
[13] M. Luwel and G. Severne, Astron. Astrophys. **152**, 305 (1985).
[14] C. J. Reidl, Jr. and B. N. Miller, Phys. Rev. E **48**, 4250 (1993).

# FIGURES

FIG. 1. The isothermal distribution in $\mu$ space. The number of particle is 1024. The full description of this distribution is given in Paper I

FIG. 2. The water-bag initial distribution in $\mu$ space. The number of particle is 1024. The full description of this distribution is given in Paper I

FIG. 3. The variation of the deviation from the equipartition ($\Delta(t)$) for N=64. Time is measured by the dynamical time. Two dashed lines are $\Delta(t)$ is constant and proportional to $t^{-1/2}$. The bends of the $\Delta(t)$ correspond to the microscopic and the macroscopic relaxation time.

FIG. 4. The microscopic relaxation time for the systems with various number of the particles. The doted line is $T = 2N$.

FIG. 5. The deviation of the locally averaged energies from the value of the equipartition. Individual energies are averaged over $20000 t_c$ around the given time $t$.

FIG. 6. The probability distribution of the macroscopic relaxation time of the system with $N = 64$. The relaxation times are calculated from the macroscopically same water-bag distribution but microscopically different random distribution realized by different seeds of random number generator.

FIG. 7. The macroscopic relaxation time for the systems with various number of the particles. For $N = 512$, the ensemble of 50 water-bag distributions shows no relaxation until $T = 10^6 t_c$. The statistical analysis gives the true relaxation time lies in the region restricted by the arrow in the 90 % confidence level. The dashed line is $T = 4 \times 10^4 N$.